# Incomplete Nutation Diffusion Imaging: an ultrafast, single-scan approach for diffusion mapping

Andrada Ianuş[1,2] and Noam Shemesh[1]*

[1]Champalimaud Neuroscience Programme, Champalimaud Centre for the Unknown, Lisbon, Portugal
[2]Centre for Medical Image Computing, Dept. of Computer Science, University College London, London, UK.

**Keywords**: Diffusion, Magnetic Resonance Imaging, isotropic encoding, ultrafast MRI, mean diffusivity,

Number of words in body of text: ~2800 (excluding references and acknowledgements)

Number of Figures: 5

**Running title:** Ultrafast MRI mapping of mean diffusivity

*Corresponding author
Noam Shemesh
Champalimaud Neuroscience Programme, Champalimaud Centre for the Unknown
Av. Brasilia 1400-038
Lisbon, Portugal
E-mail: noam.shemesh@neuro.fchampalimaud.org
Phone number: +351 210 480 000 ext. #4467

List of abbreviations:
dMRI – Diffusion Magnetic Resonance Imaging
MD – Mean diffusivity
SDE – Single diffusion encoding
IDE – Isotropic diffusion encoding
TR – Repetition time
INDI - Incomplete Initial Nutation Diffusion Imaging
q-MAS - Magic angle spinning of the q-vector
dfMRI – Diffusion functional MRI
UF-IDE – Ultrafast isotropic diffusion encoding
SNR – Signal to noise ratio
TE – Echo time
DTI – Diffusion tensor imaging
EPI – Echo-planar imaging

# Abstract

**Purpose:** Diffusion Magnetic Resonance Imaging (dMRI) is confounded by its long acquisition duration, thereby thwarting the detection of rapid microstructural changes, especially when diffusivity variations are accompanied by rapid changes in $T_2$. The purpose of the present study is to accelerate dMRI to a single scan acquisition, and to enable a more accurate estimation of diffusivity as function of time.

**Methods:** A general methodology termed Incomplete Initial Nutation Diffusion Imaging (INDI) capturing two diffusion contrasts in a single shot, is presented. INDI creates a longitudinal magnetization reservoir that facilitates the successive acquisition of two images separated by only a few milliseconds. INDI's theory is presented, followed by proof-of-concept ex- and in-vivo experiments at 16.4 T and 9.4 T.

**Results:** Mean diffusivities (MDs) extracted from INDI were comparable with Diffusion Tensor Imaging (DTI) and the two-shot IDE in the water phantom. As expected in the brain tissues, DTI provided lower MD than UF-IDE and IDE, but IDE and UF-IDE were in excellent agreement. Simulations are presented for identifying the regimes where INDI is most beneficial.

**Conclusions:** INDI accelerates dMRI acquisition to single-shot mode, which can be of great importance for mapping dynamic microstructural properties in-vivo without $T_2$ bias.

# Introduction

Methods enabling rapid acquisition of dynamic Magnetic Resonance Imaging (MRI) data have greatly impacted contemporary MRI. Functional MRI [1-5], hyperpolarized imaging, MR spectroscopy [6], MR Fingerprinting [7], and multidimensional NMR [8], are based on, and continuously benefit from, ultrafast acquisition schemes. By contrast, diffusion MRI (dMRI) methods, typically relying on Single-Diffusion-Encoded (SDE) schemes [9, 10], are not usually acquired dynamically, but their ability to probe micro-architectural features such as anisotropy [11], complex fiber configurations [12, 13], microscopic anisotropy [14-17] and cellular-scale dimensions [18-20] have made them widely applicable [21]. A few examples include early stroke detection [22-24], white matter orientation mapping [25], studies of neuroplasticity [26], or detection of microstructural aberrations upon disease [27, 28].

Rapid and dynamic determination of diffusion-derived metrics is however particularly important for diffusion functional MRI (dfMRI) [29, 30], a method aiming to detect neural activity through non-Blood-Oxygenation-Level-Dependent (BOLD) mechanisms. dfMRI evidenced faster activation dynamics and more localized activation foci compared with BOLD, suggesting it may be more closely correlated with underlying neural activity [29-31]. However, dfMRI's temporal resolution can be limited by the necessity to acquire at least two signals (one baseline image and one diffusion weighted image) for quantifying the apparent diffusion coefficient. To avoid excessive $T_1$ weighting and severe degradation in image quality, dfMRI measurements are typically separated by, typically, at least TR > 2-3$T_1$, imposing a practical limit on temporal resolution. Additionally, $T_2$ variations can occur on the timescale of a typical TR, potentially biasing the measurement and complicating the interpretation of dfMRI [32, 33].

Isotropic diffusion encoding (IDE)-based dMRI has been recently (re-)emerging as a valuable tool for speeding up the acquisition of a valuable rotationally-invariant parameter of the full diffusion tensor – its mean diffusivity (MD). Isotropic-encoding schemes have been proposed: Mori and van Zijl proposed the application of consecutive gradients along the laboratory x-, y- and z-gradients [34], while Topgaard used a similar diffusion encoding in a triple stimulated echo sequence [35]. De Graaf et al extended this idea to MR spectroscopy [36], and gradient waveforms were optimized to improve IDE's efficiency [37]. Other methods imparting different b-values within a single-scan (which, however, requires averaging for phase cycling) by making different coherence pathways sensitive to different b-values were also presented [38]. More recently, Eriksson et al have presented magic angle spinning of the q-vector (qMAS) – an elegant IDE framework harnessing harmonically-modulated gradient waveforms [39] or numerically optimized waveforms [40].

Here, we present a method termed Incomplete Initial Nutation Diffusion Imaging (INDI) which is designed to obtain both a baseline and a diffusion weighted image in a single shot without loss of SNR or temporal resolution. Nutation angles are tailored to keep a "fresh" longitudinal magnetization reservoir, such that it be used for consecutive measurements separated by only a few milliseconds, mitigating potential biases in MD quantification due to time-varying $T_2$. INDI's accuracy is validated in phantoms and in-vivo on a preclinical system. Simulations analyzing INDI SNR considerations and future applications vis-à-vis dfMRI, are discussed.

# Methods

The INDI method is presented in Figure 1, and its theory is presented in the Supplementary Material. INDI's mode of operation is rather simple: a fraction of the magnetization is rotated from the equilibrium position using a nutation angle $\alpha$, leaving (ideally) an equal magnetization pool unperturbed; a non-diffusion weighted image is then acquired. Immediately after this first acquisition, residual transverse magnetization is crushed, and all the unperturbed magnetization in the "reservoir" converted to transverse magnetization using a pulse angle $\beta$. An otherwise identical image to the previous excitation is acquired, but now the diffusion-weighting gradients are also applied (Figure 1). Thus, the two images required for quantifying diffusion coefficients are acquired with a separation of only a few milliseconds (see Figure 1 and Supporting Material).

All experiments involving animals were pre-approved by the institutional ethics committee. The phantom and ex-vivo experiments were performed on a Bruker Aeon Ascend 16.4 T scanner interfaced with an Avance IIIHD console and equipped with gradients capable of producing up to 3000 mT/m in all directions. In-vivo experiments were performed on a Bruker BioSpec 9.4 T scanner (Bruker, Karlsruhe, Germany) equipped with gradients capable of delivering up to 660 mT/m in all directions.

*Specimen preparation*. Doped water phantoms were prepared by gradually adding copper sulfate to a 30/70% (volumetric) $D_2O$/water, until a longitudinal relaxation time of ~200 ms was obtained. The solution was placed in a 5 mm NMR tube, which was sealed and placed in a 10 mm NMR tube filled with Fluorinert (Sigma Aldrich, Lisbon, Portugal). Brain samples (N = 3) were extracted from healthy male C57bl mice weighing ~25 g by standard intracardial PFA perfusion, followed by 12 h in a 4% PFA solution at 4°C, and placement in phosphate buffered saline (PBS) at 4°C. The brains were then soaked in a solution of PBS and 0.5M gadoterate meglumine (Dotarem) at a dilution of 1:200 (2.5 mM) for 12h [41], washed with PBS, and placed in a 10 mm NMR tube filled with Fluorinert. All samples were allowed to equilibrate with the surrounding constant temperature of 37°C.

*Scout INDI images.* An INDI "scout" sequence was acquired once per specimen with identical all diffusion gradients turned off. These scouts were used to correct INDI-derived maps (*vide-infra*).

*Water phantom experiments*. Following routine localization images and shimming, the water phantom was subject to three types of experiments: a "ground-truth" DTI, a conventional IDE MRI experiment, and the UF-IDE sequence. All experiments shared the following acquisition parameters: single-shot EPI, bandwidth = 652173 Hz, field of view (FOV) = 10 x 10 mm$^2$, matrix size 80 x 80 (partial Fourier encoding of 1.33, double sampled), leading to an in-plane resolution of 125 x 125 µm$^2$, with a slice thickness of 900 µm. The TR/TE was 1800/20 ms. UF-IDE and IDE diffusion gradient waveforms were generated according to [42] for isotropic encoding. Diffusion gradients [40] were placed before the refocusing pulse (i.e., the second waveform (Fig. 1) was nulled to mimimise echo time, which can be done since the waveforms are independent and self-refocusing) and had a duration of 7.5 ms and a b-value of 400 sec/mm$^2$. DTI experiments were performed using a Pulsed-Gradient-Spin-Echo sequence with $\Delta/\delta$ = 4/2 ms, six diffusion weighted images (b = 400 sec/mm$^2$, gradients applied in non-collinear directions) and six additional baseline images (b = 0 sec/mm$^2$) were acquired.

*Ex-vivo brain experiments*. In the brain sample, IDE and UF-IDE experiments were performed with identical acquisition parameters as described above for the water phantom, but with the following modifications: slice thickness was 650 µm (5 slices), b = 1000 sec/mm$^2$, and TR = 2500 ms; no double-sampling was employed.

*In-vivo experiment*. A male C57bl mouse weighing ~25 g was anesthetized with isoflurane (4% induction, 1-2% maintenance in 95% O$_2$) and placed in the scanner. A closed-loop circulating water system was used for temperature regulation, and respiration and rectal temperature were monitored continuously. Transmission was achieved through an 86 mm quadrature resonator, and the signal was detected by a 4-element array receive-only cryocooled coil (Bruker, Fallanden, Switzerland). The UF-IDE and IDE experiments were performed using the following common parameters: fat-suppressed single-shot EPI, bandwidth = 326087 Hz, FOV = 16 x 12 mm$^2$, matrix size 106 x 80 (partial Fourier encoding of 1.25), leading to an in-plane resolution of 150 x 150 µm$^2$; five slices were acquired, each 900 µm thick, and one single FOV saturation slice suppressing signals from the head's ventral part. TR/TE for UF-IDE and IDE were 1500/35 and 750/35 ms, respectively. Thirty-two dummy scans were applied to reach a stable magnetization steady-state. A b-value of 1000 sec/mm$^2$ was achieved via an IDE waveform duration of 13.6 and 5.7 ms before and after the refocusing pulse, respectively, with a gradient peak amplitude of 610 mT/m. Another identical UF-IDE experiment with 400 repetitions was acquired to assess potential benefits of recently-developed denoising schemes [43].

*Analysis*. Analysis in this study was performed using home-written code in Matlab® (The MathWorks Inc., Natick, MA, USA). All images were analyzed with the raw data, without any further post-processing. The full diffusion tensor was obtained from nonlinear fitting of the DTI data, and MD was calculated from the average of the eigenvalues. IDE and UF-IDE experiments provided the mean diffusivity directly from $MD_{IDE} = -\frac{1}{b}\log(\frac{S(b)}{S(b=0)})$ and $MD_{UF-IDE} = -\frac{1}{b}\log(\frac{S_2}{S_1-N_{12}})$, respectively, where $N_{12} = S_1 - S_2(G=0)$ from the scout. One in-vivo dataset was denoised slice-by-slice using random matrix theory [43], implemented in Matlab®, (window size = [8 8] voxels).

*INDI sensitivity simulations*. This analysis aims to quantify the sensitivity of INDI with its equal temporal resolution dMRI counterpart. Non-diffusion weighted INDI signals ($S_1 = S_2$) were computed through $S_{INDI} = \cos\left(\frac{\pi}{4}\right) * (1 - e^{\frac{TR}{T_1}})$ for a broad range of TRs between 0.5 and 5 s and biologically-relevant $T_1$s between 0.5 and 2.5 s. The corresponding, time-matched dMRI signals were computed as $S_{dMRI} = (1 - e^{\frac{TR}{2T_1}})$. All simulations assume that magnetization has been prepared in a steady state by dummy scans.

## Results

INDI principles were first tested on a simple doped water phantom. Assuming $T_1 \gg T_{EPI/2} + T_{spoil}$ (cf. Figure 1), $S_1$ and $S_2$(G = 0 mT/m) should ideally be identical for $\alpha = 45°$ and $\beta = 90°$; however, the two images are not exactly equal in practice (Figures 2A and 2B). $S_2$ signal intensity was typically somewhat weaker and spatially less homogeneous than $S_1$, particularly in the phantom experiments, likely due to its very short $T_1$ / long $T_2$ that can exacerbate uncrushed coherence pathways, as well as potential $B_1$ inhomogeneities, particularly for the $\pi/4$ pulse. Figure 2C shows the subtraction of the two signals, more clearly evidencing these differences (n.b., in the in-vivo experiments, the difference in these scout signals was much less pronounced, ~5%).

Figures 2D and 2E show the raw data for a particular instantiation of INDI, namely, the UF-IDE experiment, showing the attenuation of $S_2$ by diffusion weighting. Figure 2F shows the MD calculated directly from these raw images, without any correction applied. The uncorrected MD map suffers from two outstanding issues: (1) an artifactual spatial variation, unexpected for a homogeneous solution; (2) higher than expected MD values at this temperature. However, a simple subtraction of the scout image, $N_{12}$, from $S_1$, completely remedies these discrepancies: the mean diffusion coefficient map (Figure 2G) is both homogeneous across the slice, and it depicts the correct diffusion coefficient values as obtained from the gold standard DTI (Figure 2H). Figure 2I and Supporting Table S1 further quantify the distribution of diffusion coefficients within the sample. Clearly, all methods are in excellent agreement in this free diffusion scenario, although – as expected – a higher variance is observed for the UF-IDE due to its inherently lower SNR in the fully-relaxed regime.

To test the applicability of INDI in a biological system, we performed similar experiments in ex-vivo brains. Figure 3A shows mean diffusivity maps derived from UF-IDE (corrected with the $N_{12}$ scout image) and from standard IDE in a representative brain. UF-IDE and the conventional IDE result in very similar mean diffusivity maps, albeit the SNR is somewhat lower for UF-IDE in this fully-relaxed condition. Histograms from the entire brain are plotted in Figure 3B, whereas the median MD values arising from the different methods in the brain are tabulated in Supporting Table S1. The histograms are very similar for UF-IDE and IDE, as are the median MD values. The true correspondence between UF-IDE and its reference IDE, was investigated by plotting the mean diffusivity values in each voxel from the IDE experiments their UF-IDE counterparts (Figure 3C). The plots are well correlated (Pearson's ρ = 0.71) with very high significance (uncorrected p < 1E-7).

To ensure that UF-IDE can deliver robust images in-vivo with high temporal resolution, experiments were performed on a mouse with a temporal resolution of 1.5 seconds (Figure 4). The raw data (Figure 4A shows a representative slice) exemplifies that the quality of UF-IDE data is comparable with the corresponding two-shot IDE, with ~ 20% higher SNR for the former; when denoised with random matrix theory [43], the image quality becomes even better, with SNR gains up to a factor of ~2 (Figure 4A). The corresponding mean diffusivity maps extracted from these experiments are shown in Figure 4B and are of quite high quality considering the very high repetition rate. Histograms comparing the methods (Figure 4C) are significantly overlapping, and the correlation between UF-IDE and IDE (Figure 4D) is highly significant (Pearson's ρ = 0.43, uncorrected p < 1E-7).

To compare the SNR properties of INDI and its conventional dMRI counterpart, Figures 5A and 5B illustrate non-diffusion weighted signals (proportional to SNR up to a constant factor) for each method, for a broad range of TRs and biologically-relevant $T_1$s. Clearly, dMRI overperforms INDI for very long TRs; however, as TRs are decreased to 1-2 seconds the differences between the sequences' SNR becomes much less apparent. To analyze potential SNR enhancements by INDI, Figure 5C computes the ratio of $S_{INDI}/S_{dMRI}$. For the short TRs invariably required in high temporal resolution applications, the dominance of hot colors shows a marked advantage of INDI over the equivalent dMRI experiment. Quantitatively, INDI will provide SNR gains as long as TR < ~1.76$T_1$ (dashed line Figure 5C), although it should be noted that if INDI's scout images suffer signal loss in $S_2$(G=0), it will proportionally penalize SNR. Nevertheless, in our in-vivo experiments this was not an issue, and, in excellent agreement with the predictions of Figure 5C (TR of 1.5 s and $T_1$ of ~1.8 s), the non-denoised INDI acquisition indeed has an SNR gain of 1.20 to 1 when compared to IDE. Thus, INDI can be used to acquire the baseline and diffusion-weighted images milliseconds apart at least without suffering SNR loss, and potentially even with a modest SNR enhancement.

## Discussion

Dynamic changes in tissue ADC are at the core of diffusion fMRI methods aiming to map functional signals more intimately related with neural activity compared with their BOLD counterparts [30, 31, 44]. Disentangling changes in diffusion-driven metrics from changes in $T_2$ on the TR timescale could potentially improve the characterization of dynamic microstructural changes. Here, we have described INDI - a single-shot acquisition scheme with an inherent robustness against $T_2$ changes on the TR timescale. By harnessing a partial initial nutation of the magnetization to encode the baseline image, it is possible to acquire the diffusion weighted image only milliseconds later using the unperturbed magnetization reservoir. The TR is then fruitfully used to recover magnetization and reduce $T_1$ weighting. INDI's features were exemplified in a water phantom, where a single, time-independent diffusion coefficient exists, and which was accurately extracted from UF-IDE experiments. Both ex- and in-vivo brain experiments evidenced very good correspondence between IDE and UF-IDE data.

It is instructive to consider INDI's SNR regimes. For initial conditions satisfying fully-relaxed magnetization, the ideal INDI as here prescribed will incur a penalty of $\cos\left(\frac{\pi}{4}\right) = \frac{\sqrt{2}}{2} M_0$ whereas the corresponding dMRI will of course make use of the entire $M_0$. However, rapid acquisition schemes invariably entail non-fully relaxed conditions, where INDI's magnetization (assumed to be set into an initial steady-state by dummy scans) will have decayed by a factor of $\frac{\sqrt{2}}{2} * (1 - e^{-\frac{TR}{T_1}})$, while the temporally equivalent dMRI would decay by $1 - e^{-\frac{TR}{2T_1}}$; the factor of 0.5 in the latter exponential accounts for acquiring two dMRI images with identical temporal resolution as INDI. Theoretically, it can be shown that for $TR < \sim 1.76 T_1$, $S_{INDI} > S_{dMRI}$, and as shown in Figure 5C, for most biologically-relevant conditions, INDI could even entail moderate sensitivity *enhancements*.

INDI scout images also deserve some discussion vis-à-vis their temporal stability. The scout is used to normalize each pair of INDI images along a time series, thereby implicitly assuming that motion effects are negligible. However, in some applications, such as heart imaging, this assumption may be severely violated. In those cases, several scout images could be acquired in cine mode, i.e., with their cycle phase-locked to some external trigger and every INDI experiment measured along the cycle corrected with its phase-locked counterparts scout. Another alternative is to altogether forego the scout, and, while the *absolute* value of MD may be biased, its time-course may still be of significant value, as the bias should be constant assuming that $T_2$ does not vary on the millisecond timescale. Finally, it is worth mentioning that while we presented scout images with $\alpha = 45°$ and $\beta = 90°$, the difference between the images can be made even smaller if the specific values for the first and second nutation pulses are tweaked (data not shown). For example, while we showed worst-case scenarios for the tube of water, the brain's scout images differed by ~5%, which could be mitigated even further with tweaking of the nutation angles (data not shown). If a good balance between the scout's $S_1$ and $S_2$ is achieved, then the scout images are not required, and the INDI experiment can proceed without the normalization step.

Here, we focused on a specific implementation of INDI, namely the UF-IDE sequence, and demonstrated its feasibility and utility for assessing MD. However, INDI can be used with any gradient waveform such as Double-Diffusion-Encoding [14, 15, 45] or Nonuniform Oscillating-Gradient Spin-Echo [46].

In conclusion, the INDI pulse sequence was presented and revealed its capability of mapping accurate diffusion coefficients with good sensitivity and excellent temporal resolution. INDI's feasibility in preclinical settings was demonstrated, and its immunity towards rapid changes in $T_2$ are promising for future dfMRI experiments and other applications calling for rapid mapping of microstructural dynamics.

**Acknowledgements.**

NS gratefully acknowledges support from the European Research Council (ERC) under the European Union's Horizon 2020 research and innovation programme (grant agreement No. 679058 - DIRECT-fMRI), as well as under the Marie Sklodowska-Curie grant agreement No 657366. AI's work has been supported by EPSRC grants M507970, G007748, H046410, K020439, and M020533. Both authors thank Ms. Madalena Fonseca from Champalimaud Centre for the Unknown for assistance with the in-vivo experiments, and Mr. Jonas Olsen and Prof. Sune N Jespersen from Aarhus University for implementation of the denoising code.

# Figures/Tables

## INDI methodology

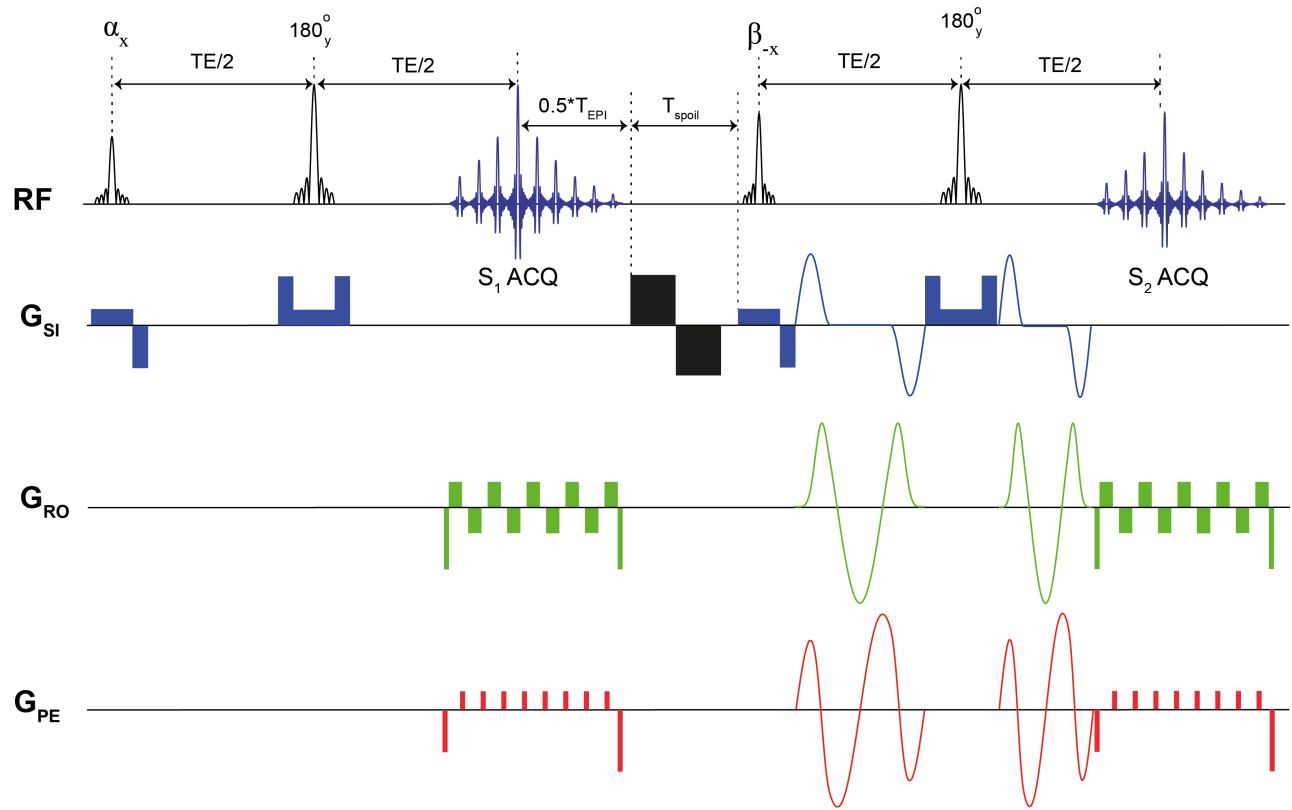

**Figure 1. General INDI methodology.** In this implementation, the sequence is furnished with isotropic diffusion encoding gradients (and hence termed UF-IDE). The sequence commences with an incomplete initial nutation, in our case $\alpha_x = \frac{\pi}{4}$. A spin-echo then proceeds, with the first acquisition providing the b = 0 s/mm² image (in our implementation, an EPI acquisition, S₁). Bipolar spoiler gradients (here shown in black) are then applied to remove the magnetization, while refocusing the residual phase to remove possible nuisance artifacts, in a similar manner to phase-rewinding in typical ultrafast imaging. Here, T$_{spoil}$ was on the order of ~10 ms. The second nutation pulse, here $\beta_{-x} = \frac{\pi}{2}$ to rotate the fresh longitudinal magnetization for the next spin-echo, which is acquired with exactly the same timing and parameters as the first echo, only the diffusion gradients are now applied (S₂). Here, we focus on obtaining the mean diffusivity, and hence we apply IDE gradient waveforms. The resulting UF-IDE sequence thus provides both baseline and diffusion weighted images within $2TE + T_{EPI}/2 + T_{spoil}$.

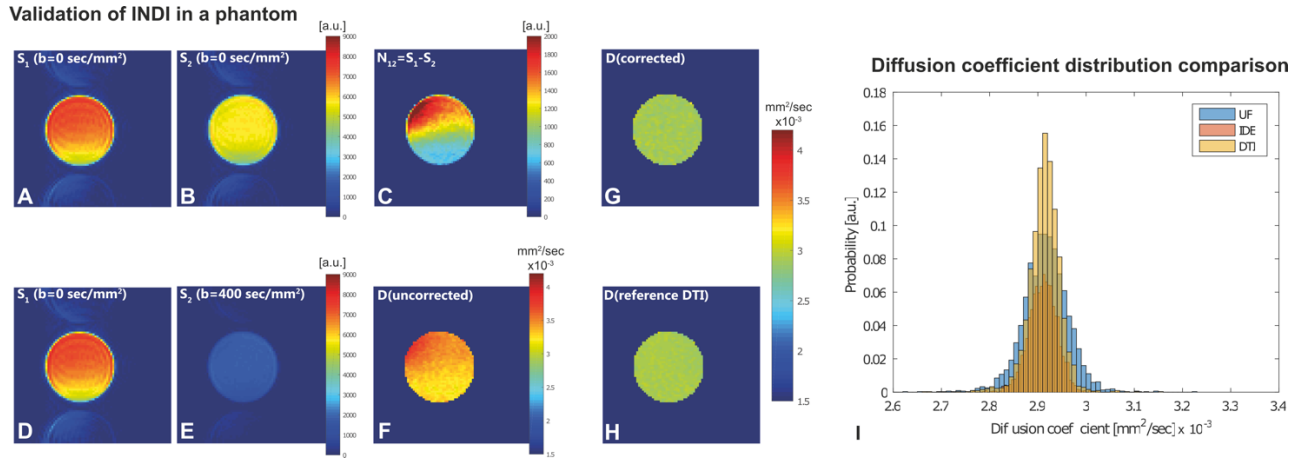

**Figure 2. Experimental validation of INDI in a phantom. (A-B)** Raw data for the scout INDI image, representing $S_1$ and $S_2$ in the absence of diffusion weighting (ideally, $S_1 = S_2$). **(C)** The difference image, $N_{12}$, clearly shows that the echoes are not ideally matched. **(D-E)** Raw data for INDI (specifically, UF-IDE). The signal in (E) is significantly attenuated by diffusion. **(F)** Mean diffusivity derived directly from the images in (D-E). The map is inhomogeneous and the diffusion coefficient is larger than expected. **(G)** Mean diffusivity calculated using a correction from the scout image, showing a homogenous image of the tube, as expected. **(H)** Ground-truth mean diffusivity from DTI. Note that there is an excellent agreement between the maps in (G) (single shot experiment) and (H) (12 different experiments separated by a single TR for every image acquired). **(I)** The UF-IDE, IDE, and DTI histograms are clearly overlapping, suggesting excellent agreement between the methods and noisier data for UF-IDE as expected at the fully relaxed condition.

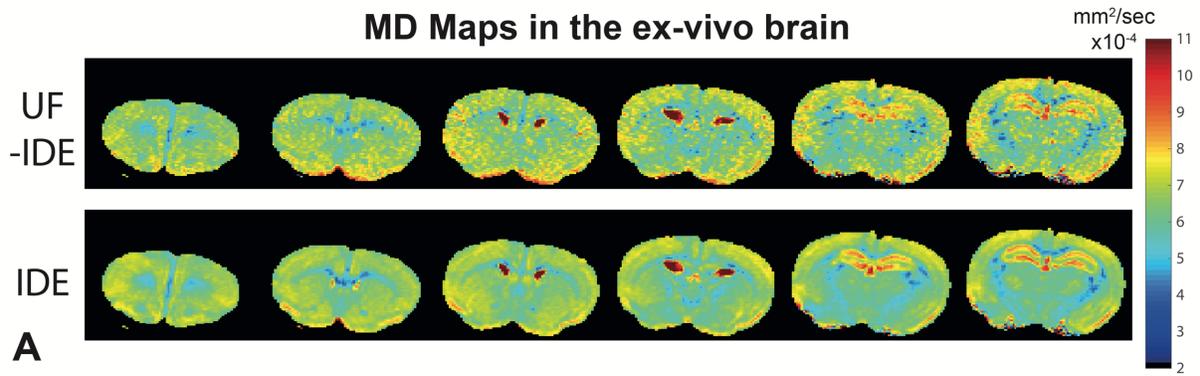

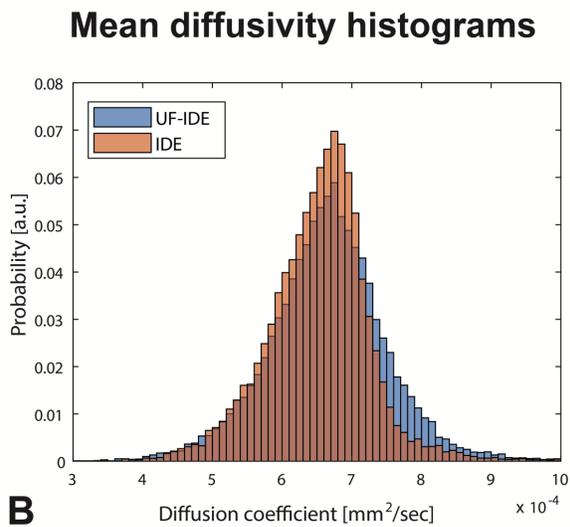
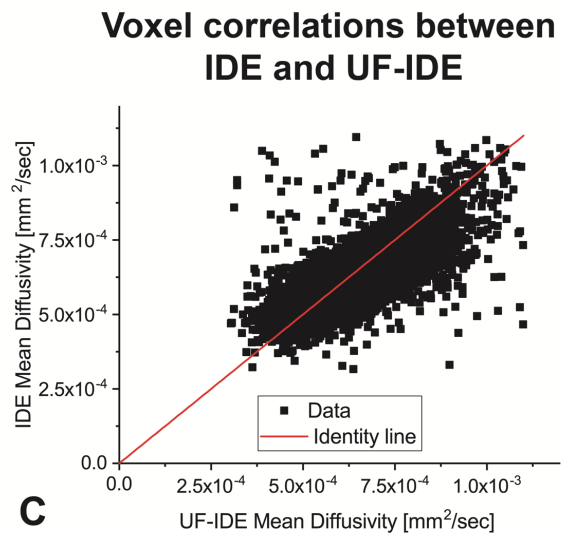

**Figure 3. Validation of INDI in ex-vivo brain. (A)** Mean diffusivity maps from UF-IDE and IDE, showing comparable MD for UF-IDE and IDE. **(B)** Histogram analysis shows very similar distribution of MD for UF-IDE and IDE. **(C)** A correlation plot for UF-IDE and IDE shows very good correspondence between the voxels acquired with different methods. All brain (but not surrounding) voxels were pooled together for both panels.

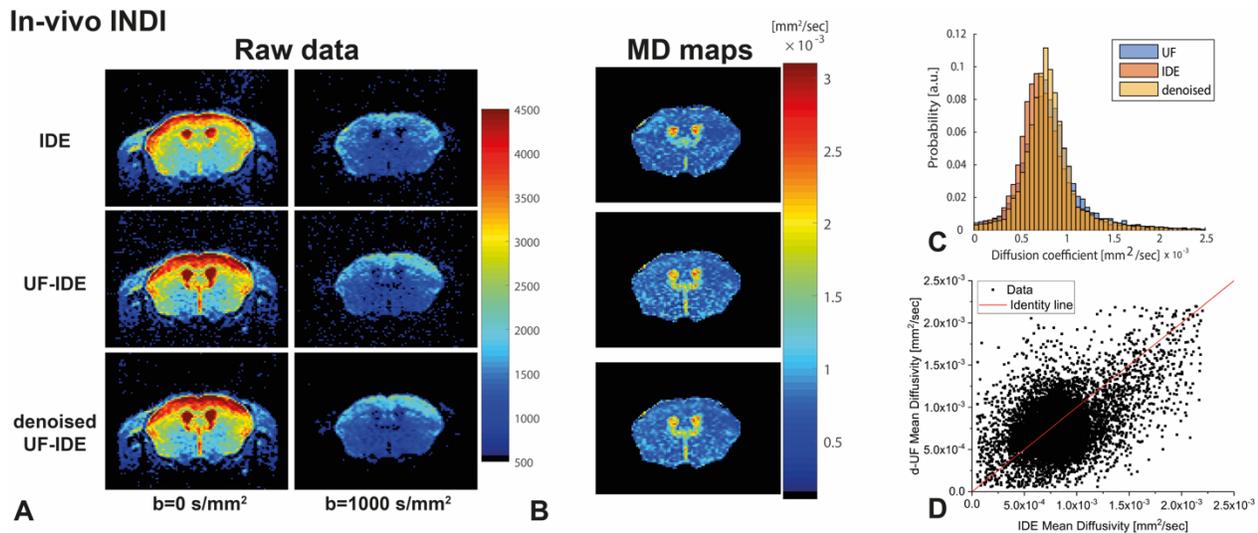

**Figure 4. INDI in-vivo. (A)** Raw data from a representative slice of the mouse brain acquired at 9.4T, for conventional IDE (acquired with TR = 750 ms), UF-IDE, and denoised IDE (d-IDE), both acquired with TR=1500 ms, but having the same temporal resolution as the conventional IDE. Excellent image quality was observed. **(B)** Corresponding MD maps from all slices. The single-shot experiments, especially once denoised, are of good quality. **(C)** Histogram distributions for the different methods for all brain (but not surrounding) voxels. The methods provide nearly identical distributions. **(D)** Correlation analysis of IDE and d-UF IDE reveals a good correlation between the methods.

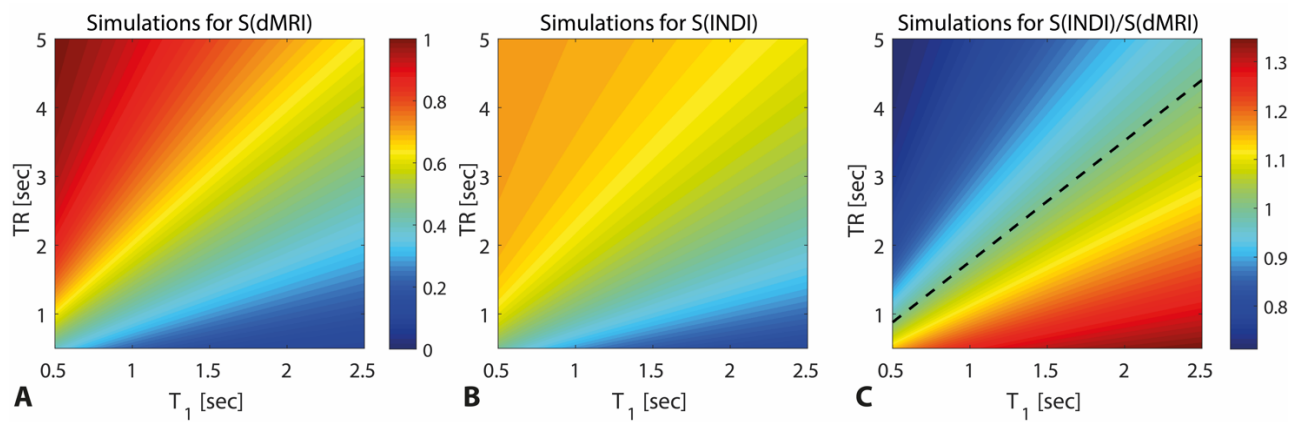

**Figure 5. SNR analysis for INDI and conventional dMRI.** Simulated signals for **(A)** conventional dMRI and **(B)** INDI over a wide range of practical TR values and $T_1$s values typical for biological tissues for field strengths between 1 and 16.4 T. Assuming constant noise with a standard deviation of one, the SNR profile of the two acquisitions is proportional to the signal maps. **(C)** INDI/dMRI signal ratios. The dashed line shows the point where dMRI and INDI have theoretically the same SNR, given that the two required images for each method are acquired with the same temporal resolution. INDI has a significant advantage when TR < ~1.76*$T_1$.

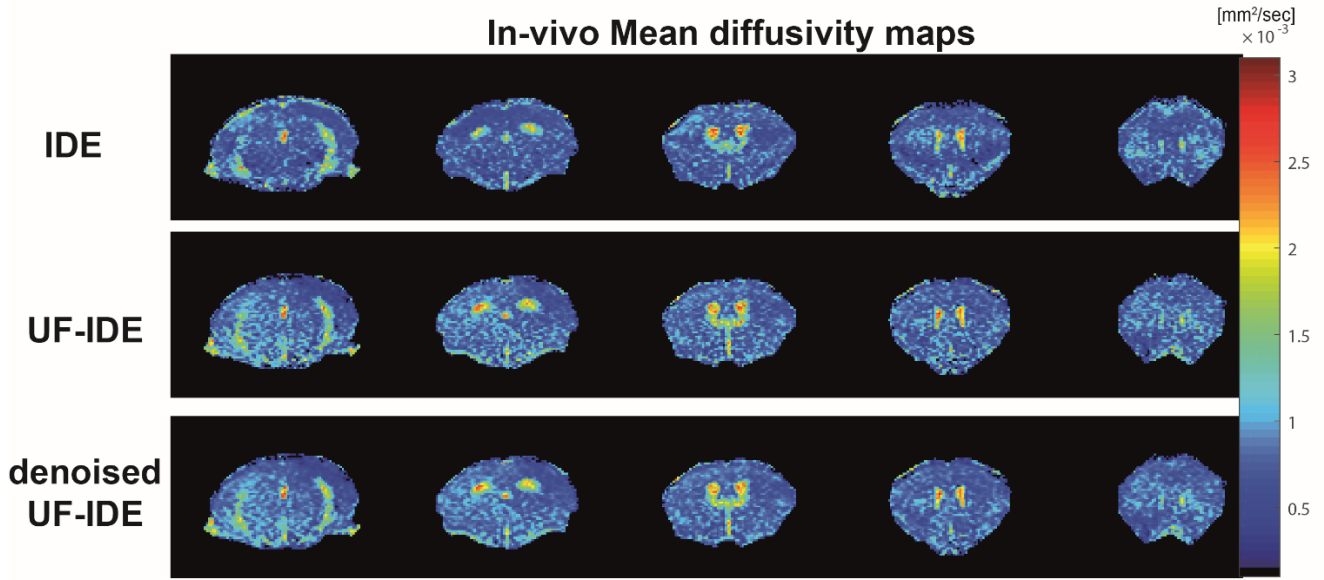

**Supporting Figure S1. INDI maps in-vivo for all slices.** Top to bottom rows represent IDE, UF-IDE, and the denoised UD-IDE mean diffusivity maps.

**Supporting Table S1**. Median and interquartile ranges for the mean diffusivity extracted from the different methods.

| Sample/Method | UF-IDE[a] | IDE[b] | DTI[c] |
|---|---|---|---|
| Water phantom | 2.92±0.06 | 2.91±0.04 | 2.92±0.04 |
| Ex-vivo Brain | 0.67±0.10 | 0.66±0.08 | - |

[a]Single shot
[b]Required two images separated by a single TR
[c]Measured with twelve images, each separated by TR